# Comparison of path following in ships using modern and traditional controllers

Sanjeev Kumar Ramkumar Sudha[1], Md Shadab Alam[2], Bindusara Reddy*[3] and Abhilash Sharma Somayajula[4]

**ABSTRACT**

Vessel navigation is difficult in restricted waterways and in the presence of static and dynamic obstacles. This difficulty can be attributed to the high-level decisions taken by humans during these maneuvers, which is evident from the fact that 85% of the reported marine accidents are traced back to human errors. Artificial intelligence-based methods offer us a way to eliminate human intervention in vessel navigation. Newer methods like Deep Reinforcement Learning (DRL) can optimize multiple objectives like path following and collision avoidance at the same time while being computationally cheaper to implement in comparison to traditional approaches. Before addressing the challenge of collision avoidance along with path following, the performance of DRL-based controllers on the path following task alone must be established. Therefore, this study trains a DRL agent using Proximal Policy Optimization (PPO) algorithm and tests it against a traditional PD controller guided by an Integral Line of Sight (ILOS) guidance system. The Krisco Container Ship (KCS) is chosen to test the different controllers. The ship dynamics are mathematically simulated using the Maneuvering Modelling Group (MMG) model developed by the Japanese. The simulation environment is used to train the deep reinforcement learning-based controller and is also used to tune the gains of the traditional PD controller. The effectiveness of the controllers in the presence of wind is also investigated.

# 1. INTRODUCTION

Shipping lies at the heart of globalization, the backbone of the modern economy. Around 80% by volume of international trade in goods is carried by sea, which has led to more traffic and higher collision risk. Despite modern ships using radar and Automatic Identification Systems (AIS) to obtain traffic data, excessive information has caused poor decision-making (Tsou et al. 2010). Autonomous ships are the modern solution to tackle such problems like safety, fuel efficiency, and vessel operations. Recent IMO regulations have pushed designers and owners to consider autonomous ships, and the Yara Birkeland project has shown success in reducing diesel vehicle trips. Human error is a major contributor to marine accidents, posing a threat to human lives, the environment, and causing significant financial losses (Baker and McCafferty 2005). The recent Ever Given incident in the Suez Canal, 2021 is just one such example, where the vessel could not maintain its path under the influence of strong winds. AI advancements, specifically in reinforcement learning, offer new solutions for ship trajectory tracking (Deraj et al. 2023) and collision avoidance without human intervention, which is crucial to reduce the likelihood of incidents.

[1] NTNU; 0000-0001-7442-3696; sanjeev.k.r.sudha@ntnu.no
[2] IIT Madras; 0000-0001-9184-9963; oe21s007@smail.iitm.ac.in
[3] IIT Madras; 0000-0001-7306-0686; na20b013@smail.iitm.ac.in
[4] IIT Madras; 0000-0002-5654-4627; abhilash@iitm.ac.in

Conventional autopilots typically use line of sight (LOS) guidance systems along with proportional-integral-derivative (PID) control systems to track waypoints for path following (Lekkas and Fossen 2012, Moreira et al. 2007). However, there is a growing interest in exploring AI-based control strategies as they have shown promising results in various applications such as active heave compensation (Zinage and Somayajula 2021, Zinage and Somayajula 2020) and dynamic positioning system (DPS) (Lee et al. 2020). Some contemporary studies have also begun to look into RL methods for path following and trajectory tracking. (Wang et al. 2019) used the Q-learning algorithm for path planning of an autonomous ship. (Sivaraj et al. 2022) used Deep Q Network (DQN) for heading control and path following of a KVLCC2 ship in calm water and waves. Studies have shown that the DRL can be employed to avoid obstacles and follow COLREGs (International Regulations for Preventing Collisions at Sea) (Heiberg et al 2022, Zhao et. al 2019). (Larsen et al. 2021) conducted a comparison of various deep reinforcement learning algorithms to assess their effectiveness in collision avoidance and waypoint tracking for ships. In addition, they evaluated their model in three distinct simulated real-world environments that were based on AIS data. (Shen et al 2019) developed and trained a Deep Q-Network (DQN) model for ship collision avoidance. They conducted experiments with three ships deployed simultaneously and studied the performance of the model in avoiding collisions and tracking waypoints. (Havenstrom et. al 2021) achieved collision avoidance and 3D path following for a 6-DOF underwater vehicle with help of the Proximal Policy Optimization (PPO) algorithm. (Zhou et. al 2019) implemented path planning for a single USV and USV formation whilst also avoiding stationary obstacles using deep Q-network (DQN).

There exist several studies that have used various planning algorithms to implement collision avoidance and path planning in surface vessels. (Chun et. al 2021) applied the PPO algorithm to implement COLREGs, compared it against the A* algorithm and observed that the maximum Collision Risk (CR) was significantly reduced with the PPO algorithm. (Woo and Kim 2020) utilized a Semi Markov Decision Process (SMDP) to model a USV to comply with COLREGs and suggested a technique where an USV can perceive its surroundings by utilizing a grid map representation. (Lyu and Yin 2019) implemented COLREGs compliant behaviour by utilizing Artificial Potential Field (APF), which includes a repulsion potential field function and corresponding virtual forces, to effectively address the collision avoidance problem. (Chen et al. 2017) used barrier function for obstacle avoidance, and compared it with a potential field method and the Hamilton-Jacobi method (Takei et al. 2010). (Xu et. al 2022) used path planning and dynamic collision avoidance (PPDC) algorithm to ensure safe navigation of USVs. (Zhao et. al 2019) employed the PPO algorithm for path following while complying with COLREGs and showed that it performed better when compared with a PID controller.

Most of the research on Deep Reinforcement Learning (DRL) has been centered on employing it for obstacle avoidance and path planning, with only a limited number of studies comparing its effectiveness to conventional methods for path following. Nonetheless, a great deal of work is still required before RL-based controllers can be suitably integrated into real-world scenarios, as simulations often oversimplify the real-world dynamics and neglect environmental forces. Therefore, the objective of this study is to develop a Proximal Policy Optimization (PPO) based controller for a ship's path following task, which is characterized by a strongly non-linear model, and to assess its performance in the presence of significant environmental forces.

The rest of the paper is organized as follows. Sec. 2 deals with the dynamics of the ship. Sec. 3 introduces the basics of reinforcement learning and the PPO algorithm. Sec.4 illustrates on how the algorithm mentioned in Sec. 3.1 is applied to the problem of ship path following problem. Sec. 5 describes the results obtained and further analyses the performance of the PPO agent by evaluating various maneuvers in calm water and in presence of winds. Sec. 6 compares the performance of PPO agent against a traditional PD controller. Finally, Sec. 7 summarises the results and discussion of this study.

## 2. SHIP DYNAMIC MODEL

The KCS vessel is chosen to run numerical simulations and test the control algorithm used in this study. The MMG (Maneuvering Modelling Group) model (Yoshimura and Masumoto 2012, Yasukawa and Yoshimura 2015) is used to mathematically model the ship dynamics. The specifics of the KCS vessel and description of its dynamics are provided in (Deraj et al. 2023). Ship maneuvering in surge, sway, and yaw motions are solved using 3-DOF nonlinear equations of motion, after being non-dimensionalized according to the prime-II system of normalization given in (Fossen 2011).

## 3. REINFORCEMENT LEARNING

Reinforcement learning problems involve training intelligent agents that interact with a complex environment and learn to take optimal decisions such that a numerical reward is maximized. In such problems, the environment is modelled as a Markov decision process (MDP). In an MDP, the agent interacts with the environment by choosing an action *a* based on the observations of the state *s* and receives a numerical reward *r* in response, while transitioning to the next state *s'*. The policy $\pi(s)$ refers to the method by which the agent selects specific actions. The primary objective of the agent is to maximize its long-term reward, usually the episodic return, which is the sum of rewards in an episode.

### 3.1 Proximal Policy Optimization (PPO)

PPO (Schulman et al. 2017) algorithm is an actor critic, on-policy DRL algorithm that is based on trust region policy optimization (TRPO) (Schulman et al. 2015). It is an on-policy algorithm, meaning the agent's current policy, and the behavioural policy which is used to collect data for training are the same. In an iteration, a set of transitions (*s, a, r, s'*) are collected by running a certain number of episodes using the current policy before it is updated.

## 4. IMPLEMENTATION OF PPO ALGORITHM FOR SHIP NAVIGATION

### 4.1 Observation State Space

The PPO algorithm utilizes two neural networks, namely actor and critic networks, which use the state as input information. As a result, all states must be observable. The state is comprised of four elements, which is a noteworthy characteristic of this study as the agent was able to reach the destination with a limited number of parameters compared to other published works. The state vector comprises of:

$$s_t = [d_c, \chi_e, d_{wp}, r] \quad (1)$$

where, $d_c$ is the cross-track error, $\chi_e$ denotes the course angle error, $d_{wp}$ is the distance to destination and *r* is the yaw rate.

## 4.2 Action space

The commanded rudder angle, $\delta_c$ is the action that the agent can choose at a given time step and its value lies in the range $\delta_c \in [-35°, 35°]$.

## 4.3 Reward Structure

In Reinforcement Learning (RL), the agent aims to optimize a numerical signal known as the reward, which is provided by the environment as a response to the agent's actions. The reward serves as a feedback mechanism for the agent, informing it of its performance in a particular state and guiding its future actions towards its objective. In DRL, the agent is a neural network that receives the environment's observations as input and generates actions as output. The network's parameters are adjusted by an algorithm based on the rewards it receives from the environment. Rewards play a pivotal role in the DRL agent's learning process, as it tries out various actions and observes the corresponding rewards. Positive rewards reinforce the agent's decision-making process, while negative rewards discourage it from repeating certain actions. As the agent gains experience, it identifies which actions result in the highest rewards and updates its policy to maximize future rewards. Therefore, rewards offer a crucial feedback signal for the DRL agent to learn and enhance its decision-making abilities. The rewards obtained at any intermediate time step are given by:

$$
\begin{aligned}
r_1 &= 2 \exp\left(\frac{-d_c^2}{12.5}\right) - 1 \\
r_2 &= 1.3 \exp(-10|\chi_e|) - 0.3 \\
r_3 &= \frac{-d_{wp}}{4}
\end{aligned}
\quad (2)
$$

where $r_1$ is the reward associated with cross track error, $r_2$ is the reward associated with course angle error. Finally, to make overall reward negative, we introduced reward $r_3$ that depends on the distance between ship current position and the destination waypoint.

The reward at time step *t* is denoted by $r_t$ and is given as the sum of the rewards from each component at that time step.

$$
\begin{aligned}
r_t &= r_1 + r_2 + r_3 \\
R &= \sum_0^t r_t
\end{aligned}
\quad (3)
$$

In eq. (3), $r_t$ is the reward at time step *t* and *R* is the episode return, which is the cumulative sum of rewards obtained at each time step. Finally, a reward of *+100* is given to agent to the agent if it successfully reaches the destination.

## 4.4 Training Process

In each episode, the ship begins at the origin (initial waypoint), oriented in positive x-direction (heading angle $\psi$= 0) and having an initial velocity in surge direction (*u*) of 1 in non-dimensional form. The ship has no initial velocity in sway and yaw motion. The goal waypoint is chosen randomly between *8L* to *28L* from origin, where *L* is the ship's length between perpendiculars. The direction is chosen uniformly between 0 to $2\pi$.

The horizon for an episode is set to be 160 time-steps. A *0.5L* tolerance is specified on the destination waypoint, and the episode is considered successful if the ship can enter this region. Another condition is established to determine whether the agent is still capable of reaching the destination point or if the episode is a failure and the ship is just loitering around. The termination condition is as follows:

$$\vec{v_1} \cdot \vec{v_2} < 0 \text{ and } \vec{U} \cdot \vec{v_2} < 0 \tag{4}$$

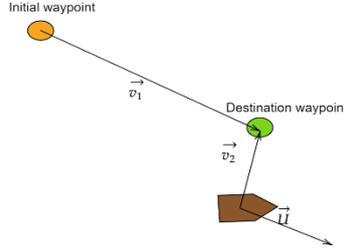

Figure 1: Termination condition of the agent

### 4.5 Hyperparameters of the networks

The agent is tuned by experimenting with different values for the actor, critic networks and other hyperparameters. The actor network as well as critic network both uses 2 hidden layers with *tanh* activation in the hidden layers. As PPO is an on-policy algorithm, data is collected for over 50 episodes in each iteration before updating the actor and value networks over a certain number of epochs. The best set of hyperparameters used for training are given in Table I. Fig. 2 shows the plot of the average returns in every iteration during training. The policy at 60 iterations is chosen to strike a balance between underfitting and overfitting.

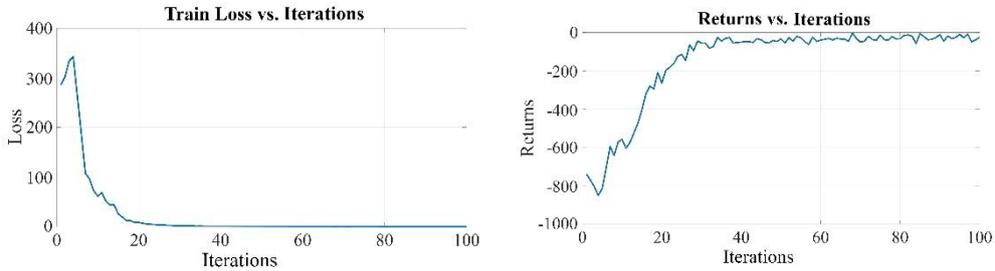

Figure 2: Training Loss and Returns for the model

TABLE I: Hyperparameters for PPO Model

| Hyperparameter | Value |
| --- | --- |
| Initial Learning rate | 0.001 |
| Decay Steps | 3000 |
| Decay Rate | 0.5 |
| Actor Hidden Layers | 128,128 |
| Critic Hidden Layers | 128,128 |
| Discount Factors | 0.96 |
| Clip Ratio | 0.2 |
| Lambda | 0.95 |
| Entropy Regularisation | 0.2 |
| Training epochs | 10 |
| Episodes per iteration | 50 |
| Number of iterations | 100 |

## 5. RESULTS

A three-step evaluation process is used here to assess the performance of the PPO agent. First the agent's ability to track waypoints in different situations is evaluated selecting destination waypoint in each quadrant. Next, the agent is directed to follow complex path discretized into number of waypoints. Finally, wind is introduced in the environment and agent is tested under the influence of wind forces.

### 5.1 Waypoint Tracking

The performance of the PPO agent is analyzed by giving destination waypoint in all four quadrants namely (*10L, 10L*), (*-10L, 10L*), (*10L, -10L*) and (*-10L, -10L*). The starting point of the vessel is chosen at origin with heading angle ($\psi$) as 0 and a surge velocity of *1U*. Fig. 3 shows that the trained model is successfully able to reach the goal waypoint in all four quadrants.

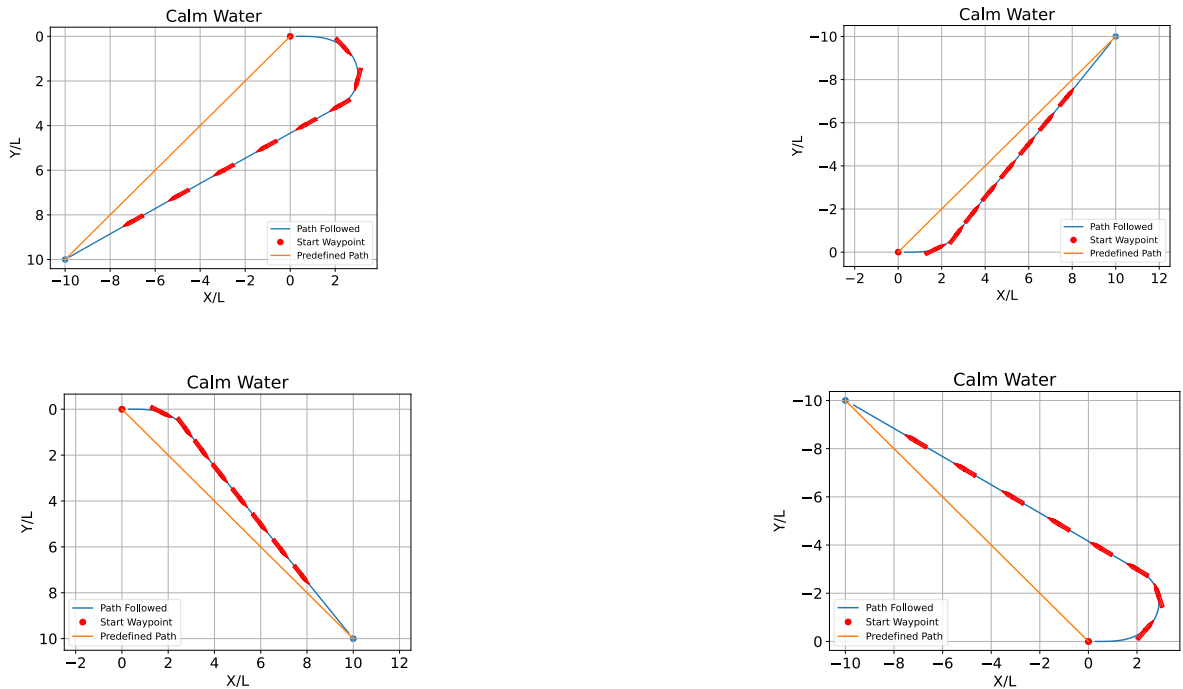

Figure 3: Single waypoint tracking

### 5.2 Path following through waypoint tracking

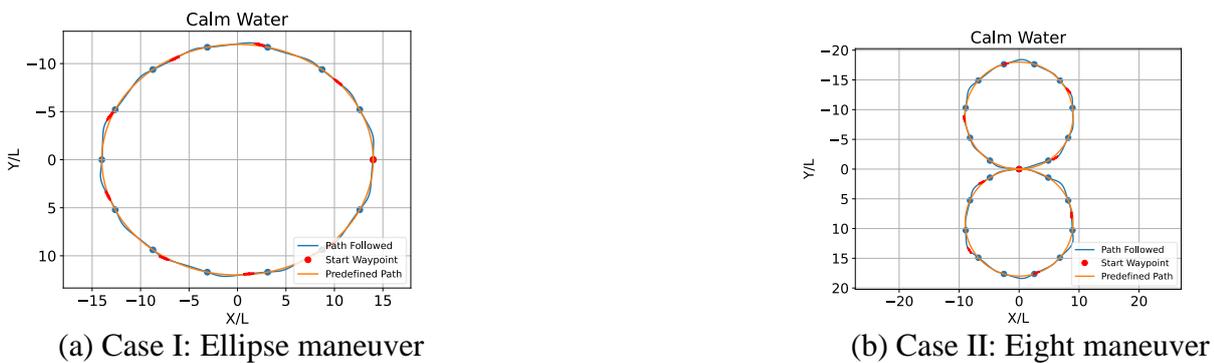

(a) Case I: Ellipse maneuver  (b) Case II: Eight maneuver

Figure 4: Path Following cases

As the model successfully tracks waypoints in all four quadrants, it is capable of navigating more complex paths. The trajectory of the model for various shapes is illustrated in Fig. 4. Fig. 4(a) demonstrates the agent's capacity to follow an elliptical path. The ellipse, with major and minor axes of lengths *28L* and *24L*, respectively, is discretized into 15 waypoints. The vessel begins at $(x, y) = (14L, 0)$, with an initial heading of $\psi = -\pi/2$ along the negative Y-axis. A path in the shape of the numeral "8" is defined and discretized into 23 waypoints. Fig. 4(b) indicates that the ship initiates the eight maneuver at the origin with a heading of $\psi = 0$ and successfully completes the lower circle of radius *9L* before finishing the upper circle of the same radius.

## 5.3 Performance with wind forces

Wind forces and moments are modelled and added as described in (Deraj et. al 2023). Fig. 5 displays two specific cases that were examined, each with different wind speeds and directions. It can be observed that the the agent is able to get the ship to follow the desired trajectory even in the presence of strong winds with a wind speed that is six times the ship's operating design velocity.

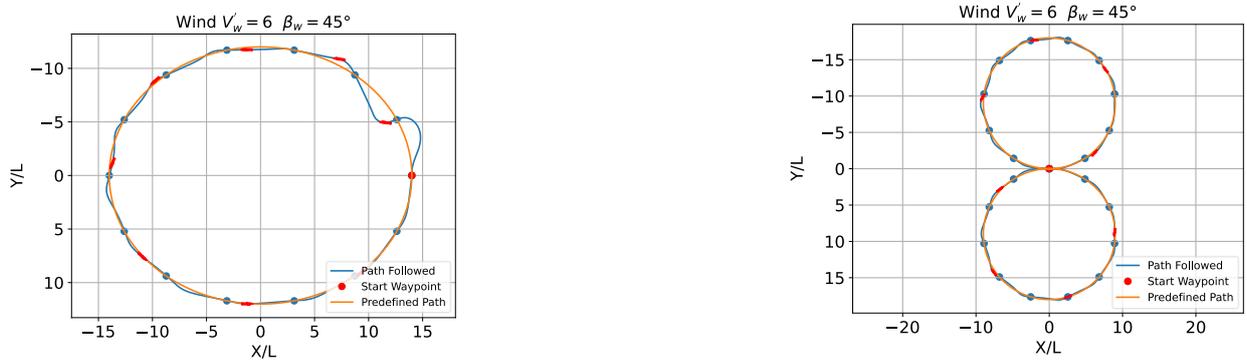

Figure 5: Path following in presence of constant and uniform wind

## 6. COMPARISON WITH PD CONTROLLER

This section compares the path following ability of the PPO agent with that of a Proportional Derivative (PD) controller. The desired heading angle $\psi_d$ is obtained using the integral line of sight (ILOS) guidance law (Fossen 2021). The error $e$ is calculated as the difference between the current heading angle $\psi$ and the reference value $\psi_d$ and is expressed in eq. 6(a). Meanwhile, the derivative of the error with respect to time is represented by eq. 6(b).

$$e = \psi_d - \psi \qquad (6(a))$$

$$\dot{e} = \dot{\psi}_d - \dot{\psi} = -r \qquad (6(b))$$

Thus, the proportional derivative (PD) control law can be expressed as

$$\delta_c = K_p e + K_d \dot{e} = K_p(\psi_d - \psi) - K_d r \qquad (7)$$

where $\delta_c$ is the commanded rudder angle. $K_p$ is the proportional gain and $K_d$ is the derivative gain of the controller. The controller gains are tuned to minimize the tracking error. The values of the proportional and derivative gains used are $K_d = 4.0$ and $K_p = 2.0$ respectively. Fig. 6 shows a comparison of trajectories between the PPO controller and the PD controller, for square and eight maneuvers. It is observed that root mean square (RMS) cross-track error with the PPO based controller is *45.5%* smaller than that of the PD controller for a square trajectory. Similarly, the RMS of cross track error is calculated to be *19.8%* smaller in case of an eight trajectory which consists of 20 waypoints with circles of radius *6L*. Fig. 7 shows the variation of rudder angle for both the controllers while tracking the eight trajectory. Although the cross track error is lesser with the PPO controller, the controller effort is more, as seen in Fig. 7.

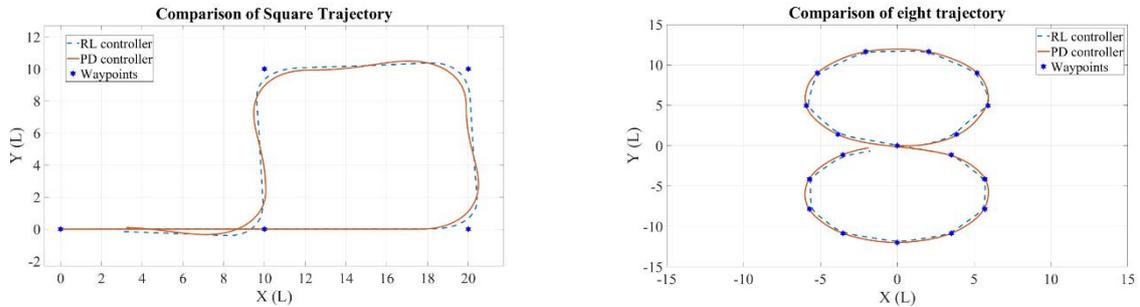

Figure 6: Comparison of path traversed by the model in PD controller and PPO controller

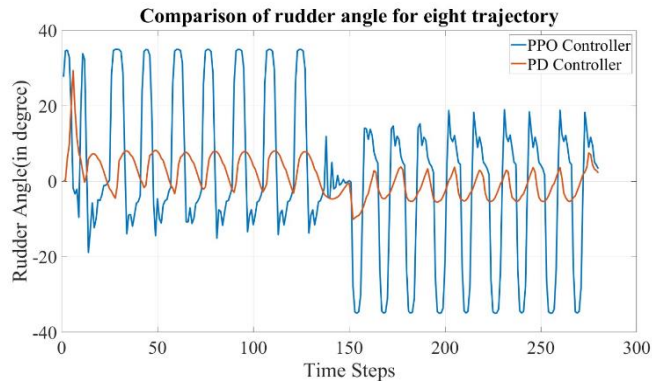

Figure 7: Rudder angle variation during the eight maneuver

## 7. CONCLUSION

A DRL-based controller has been utilized in this study for the purpose of path following of a ship via waypoints. To achieve this task, a PPO agent was trained and given rewards that were associated with cross-track error, course angle error, and distance to the target waypoint. The PPO agent successfully demonstrated its ability to follow destination waypoints in still waters, as well as complex paths such as ellipse and numeral eight maneuvers that were discretized by waypoints. Further, it is found to be as effective for trajectory tracking as a PD controller, which is standardly used for this task. It is also observed that the controller effort in case of the PPO controller is significantly higher as compared with the PD controller, which can be improved with more hyperparameter tuning of the RL agent.

Moving forward, the current DRL framework shall be improved to include obstacle and collision avoidance, as well as adherence to COLREGs, allowing for the optimization of multiple objectives. This will result in the creation of a policy that is both computationally efficient and allows implementation in real-time. Multi-objective control structures will be experimentally tested to gain a better understanding of traditional and modern controller's subtleties. Future studies will also explore the effectiveness of the controllers under other environmental disturbances such as waves and currents.

## 8. ACKNOWLEDGEMENT

This work was partially funded by the Science and Engineering Research Board (SERB) India - SERB Grant CRG/2020/003093 and New Faculty Seed Grant of IIT Madras. This work is also supported through the proposed {Center of Excellence for Marine Autonomous Systems (CMAS), IIT Madras} setup under the Institute of Eminence Scheme of Government of India.